\documentclass{INTERSPEECH2023}
\usepackage{caption}
\usepackage{subcaption}
\usepackage{multirow}
\usepackage{tabularx}
\usepackage{etoolbox}
\newcolumntype{L}[1]{>{\raggedright\let\newline\\\arraybackslash\hspace{0pt}}m{#1}}
\newcolumntype{C}[1]{>{\centering\let\newline\\\arraybackslash\hspace{0pt}}m{#1}}
\newcolumntype{R}[1]{>{\raggedleft\let\newline\\\arraybackslash\hspace{0pt}}m{#1}}

\interspeechcameraready

\title{DC CoMix TTS: An End-to-End Expressive TTS with Discrete Code Collaborated with Mixer  }
\name{Yerin Choi, Myoung-Wan Koo}
\address{
  Department of Artificial Intelligence, Sogang University, Republic of Korea}
\email{lakahaga@sogang.ac.kr, mwkoo@sogang.ac.kr}

\begin{document}

\maketitle
 
\begin{abstract}
Despite the huge successes made in neutral TTS, content-leakage remains a challenge. In this paper, we propose a new input representation and simple architecture to achieve improved prosody modeling. Inspired by the recent success in the use of discrete code in TTS, we introduce discrete code to the input of the reference encoder. Specifically, we leverage the vector quantizer from the audio compression model to exploit the diverse acoustic information it has already been trained on. In addition, we apply the modified MLP-Mixer to the reference encoder, making the architecture lighter. As a result, we train the prosody transfer TTS in an end-to-end manner. We prove the effectiveness of our method through both subjective and objective evaluations. We demonstrate that the reference encoder learns better speaker-independent prosody when discrete code is utilized as input in the experiments. In addition, we obtain comparable results even when fewer parameters are inputted.
\end{abstract}

\section{Introduction}

Text-to-Speech (TTS) has made remarkable progress with the use of deep neural networks. From cascaded \cite{Wang2017TacotronTE,Ren2019FastSpeechFR,Ren2020FastSpeech2F,ren2021portaspeech} to end-to-end models \cite{kim2021conditional,lim22_interspeech}, recently proposed systems have achieved both human-like quality and fast inference speeds. Various generative models like Flow and DDPM \cite{ho2020denoising} have improved TTS further \cite{shih2021rad,lam2022bddm,DBLP:conf/ijcai/HuangL0S00Z22}. However, when it comes to expressive TTS, which leverages additional prosody information to synthesize expressive speech, has not achieved as high of success compared to general TTS. One subfield of expressive TTS is Prosody Transfer, which learns to encode fine-grained prosody from speech and transfer it to synthesized speech. In prosody transfer, some challenges, such as the content leakage problem, still need to be solved. Our primary purpose is to explore better prosody modeling that can resolve the issues in prosody transfer TTS.

Prosody Transfer TTS, also called unsupervised expressive TTS, intakes the reference speech and input text and synthesizes the speech with desired prosody. The reference encoder uses speech to learn prosody representation and extracts utterance-level style embedding from the given reference speech in inference \cite{wang2018style}. The synthesized audio often has unwanted contents from reference speech (i.e., the content leakage problem). Prosody is hard to learn because it is compound information that comprises several features (e.g., rhythm, timbre, etc.). Recent approaches adopt mutual information minimization in training the reference encoder so that the style embedding can only hold acoustic information from the reference speech \cite{hu2020unsupervised,yi2022prosodyspeech}. These methods often train the system module by module due to the instability of the training process. Another solution suggests using latent representation extracted from a self-supervised speech model \cite{chen2022fine}. This method utilizes the properties of latent representations, which are more high-level information than the Mel-spectrogram. Since self-supervised speech models are trained to learn semantic and coarse-grained acoustic features, it lacks information to generate speech \cite{Borsos2022AudioLMAL}.

Recently, many studies have found that discrete code is a good representation of generation tasks. From the fields of TTS \cite{liu22c_interspeech,du22b_interspeech,Chen2023AVQ,Wang2023NeuralCL} to speech generation \cite{Borsos2022AudioLMAL}, several studies have proved that discrete code extracted from the audio has abundant acoustic information for reconstruction. TTS models \cite{liu22c_interspeech,du22b_interspeech,Chen2023AVQ} that train on the discrete code achieved better performance than those on Mel-spectrogram.  In particular, AudioLM and VALL-E exploit neural audio compression models \cite{Zeghidour2022SoundStreamAE,Defossez2022HighFN}. They learn to predict multi-codebook codes with transformers like language models do, then use an off-the-shelf neural codec decoder to generate the waveform. We discuss more advantages of adopting the audio compression model rather than training a vector quantizer in \ref{subsection:codecinfo}.

Along with the success of neutral TTS with discrete code as an acoustic token, we propose using discrete codes extracted from the audio compression model to learn style embedding. Discrete codes from audio compression models have rich acoustic information that is sufficient for generation tasks like TTS. 
In addition, we introduced modified MLP-Mixer \cite{Tolstikhin2021MLPMixerAA} to the reference encoder, making the architecture simpler than that of GST. The modification is the same as that from MixerTTS \cite{tatanov2022mixer}, which applies depth-wise 1D convolution instead of the MLP layer. Lastly, we use VITS as the backbone TTS model, constructing an end-to-end expressive TTS model. We evaluate our proposed model using similarity MOS conducted by native English speakers. Along with human feedback, we also evaluate our system using objective metrics. Our system transfers prosody better than GST with Fastspeech2 and Hifi-GAN \cite{kong2020hifi}, preserving the desired speaker identity. We further investigate the impact of each factor of our method with both subjective and objective evaluation. We make our code and samples public.\footnote{\url{https://github.com/lakahaga/dc-comix-tts}}

\begin{figure*}[th]
  \centering
  \begin{subfigure}[t]{1.2\columnwidth}
    \centering
    \includegraphics[width=\textwidth]{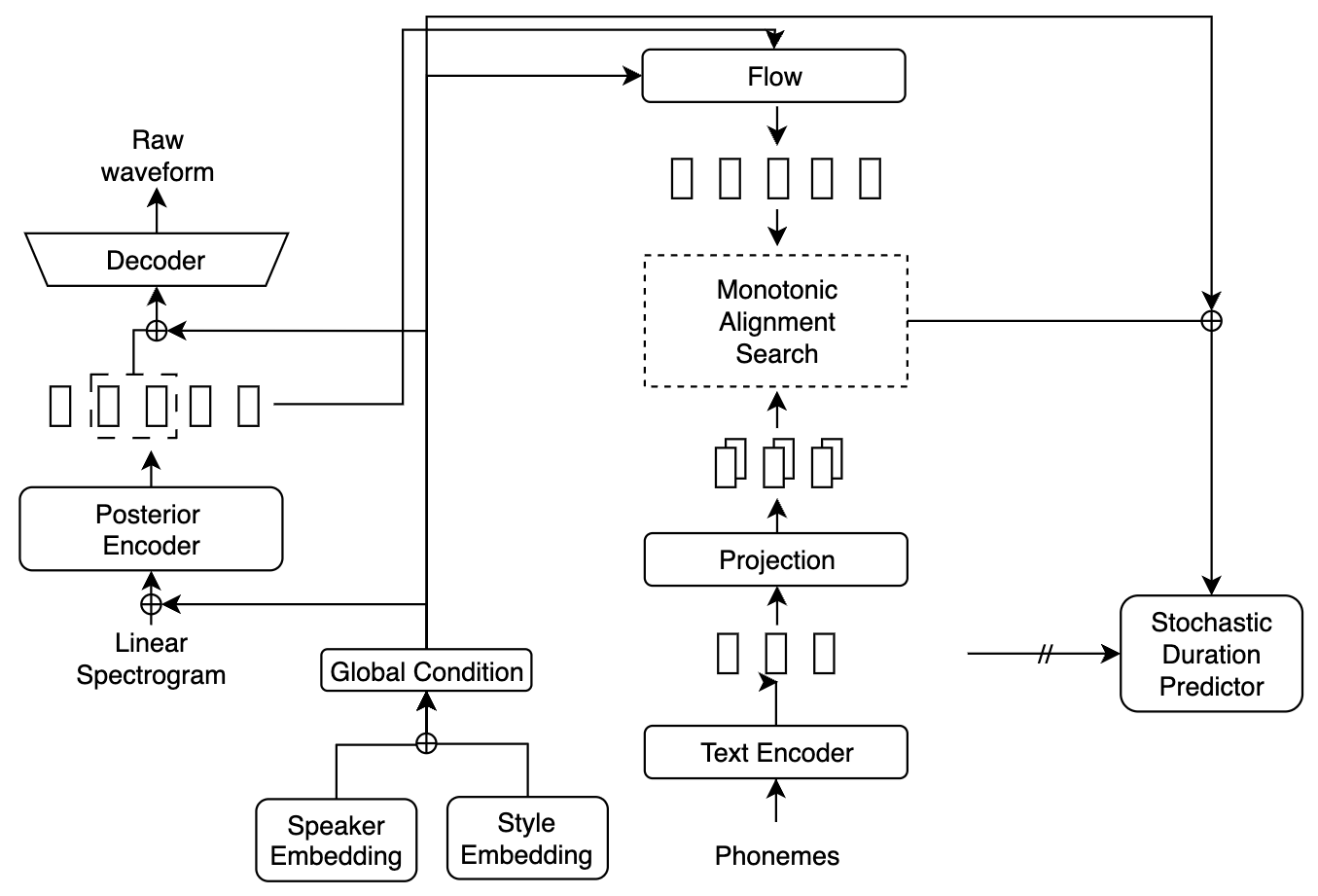}
    \caption{DC Comix TTS.}
    \label{fig:total}  
  \end{subfigure} %
  \hfill \begin{subfigure}[t]{0.8\columnwidth}
    \centering
    \includegraphics[width=\textwidth]{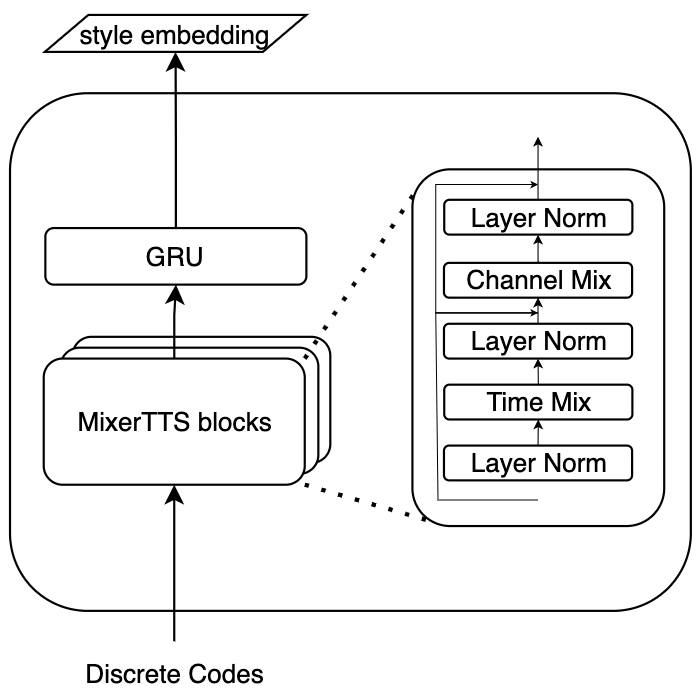}
    \caption{Reference Encoder}
    \label{fig:ref}
  \end{subfigure}
  \caption{The overall architecture of DC Comix TTS. (a) DC Comix TTS follows VITS architecture. Speaker embedding and style embedding are used as a global condition. For speaker embedding, we use learnable embedding. (b) Detailed Composition of the reference encoder. It consists of 6 layers of MixerTTS blocks and a layer of GRU.}
  \label{fig:all}
  \vspace{-1.5em}
\end{figure*}

\section{Related works}

Global Style Token \cite{wang2018style} was the first process introducing unsupervised modeling into expressive TTS. It encodes style embedding from Mel-spectrogram with multi-layer CNNs and one layer of GRU \cite{69e088c8129341ac89810907fe6b1bfe}. However, the style is not sufficiently broken down for it to be transferred to other speech using GST. It suffers from less accurate pronunciation or even wrong content. The error is due to reconstruction loss from the reference speech. The solutions to content-leakage problems can be categorized into two groups: one is adding architecture or learning objectives, and the other is changing the input representation of the reference encoder.

One suggested solution is to add an explicit module to disentangle style and content. Prosodyspeech \cite{yi2022prosodyspeech} leverages mutual information minimization when they train the reference encoder. It is trained to minimize shared information between style embedding and the corresponding content embedding. In addition, it uses cross-attention between content embedding from the given text and style exemplars from the reference speech. The authors train this model across several steps due to the instability of the training. This process of training module by module can be very tiresome since TTS models usually need numerous epochs until they converge.

Another previous way to better disentanglement style and content is to use latent representation from the pre-trained model as the input of the reference encoder. FGTransformer \cite{chen2022fine} uses wav2vec2.0 \cite{baevski2020wav2vec} embedding as a global and local style token. It distinguishes the content information of the reference speech from the synthesized speech using high-level representation. It is an effective solution for less reconstruction of its content, but it is not effective when it comes to giving sufficient information to the reference encoder to learn prosody for the style transfer.    Self-supervised speech models are trained to learn semantic and coarse-grained acoustic information (e.g., speaker identity) rather than fine-grained style information \cite{Borsos2022AudioLMAL}. Subsequently, latent representation from pre-trained models lacks the fine-grained acoustic information imperative to prosody modeling in TTS. 

Discrete code as an acoustic feature has shown promising results in TTS and speech generation. VQTTS \cite{du22b_interspeech} introduced discrete code as an intermediate representation. It showed better performance than using a pre-determined feature like Mel-spectrogram. In addition, MQTTS \cite{Chen2023AVQ} proposes multi-codebook learning to use the real-world corpus to train the TTS system. Meanwhile, AudioLM \cite{Borsos2022AudioLMAL} reached SOTA in speech generation using acoustic code extracted from the neural audio compression model. Also, VALL-E uses a similar architecture to AudioLM to introduce in-context learning in TTS. These models adopt an audio compression model. They do not train the quantizer but use it as a feature extractor.

\section{Method}

The overall architecture of our system is shown in Figure \ref{fig:total}. Our method encodes style embedding from given discrete codes of the reference speech and utilizes the style embedding as global conditioning to synthesize. The backbone TTS model is VITS, and for speaker embedding, we use the learnable look-up table.

\subsection{Discrete code for style embedding}
\label{subsection:codecinfo}

We adopt the audio compression model to address the content leakage problem and to be applicable to abundant acoustic representations. Specifically, we use Encodec\ to achieve discrete codes ${S}^{D\times{T}}$, where D is the number of codebooks in Encodec, and T is the time step. Encodec takes raw waveform $X$ as inputs and returns discrete code ${S}^{D\times{T}}$.  The Encodec quantizes audio across multiple steps, which is a process called residual vector quantization. We extract discrete code from this residual quantizer. The feature extraction process is stated in Equation \ref{eq:encodec}, where $Q_{AC}$ is the vector quantizer in Encodec. The quantizer in Encodec uses eight codebooks, so we get an 8-dimension representation. This needs much less computation than  the 80-dimension Mel-spectrogram. 
\begin{align}
  {S}^{D\times{T}} = Q_{AC}(X)
\label{eq:encodec}
\end{align}

We use an off-the-shelf audio compression model instead of a randomly initialized vector quantizer. The benefits of doing so are as follows. First, an audio compression model's learning objective is to compress and reconstruct audio. The encoder compresses the audio, and the decoder then reconstructs it. The process is trained in an end-to-end manner, so the encoder learns to encoder a low dimensional vector usable for the reconstruction step. Hence, we get a compact representation that is suitable for a generation. Second, the audio compression model is trained on an enormous and varied corpus. The Encodec model uses three corpora to train the mono-channel compression model. The total duration of the corpora is 11,629 hours. In addition, Encodec uses not only speech but also general audio. Thus, it has already learned using diverse acoustic information. We can distill this diversity by adopting Encodec without additional training. Taking advantage of this, we use Encodec as a feature extractor and feed it to the MixerTTS blocks.

\subsection{DC CoMix}

We introduce MLP-Mixer to the reference encoder to simplify the end-to-end TTS training. The architecture is shown in Figure \ref{fig:ref}. We leverage the modified MLP-Mixer blocks from MixerTTS. MixerTTS blocks comprise time-mix and channel-mix. This is because the time step is a crucial piece of information in speech. After the MixerTTS blocks, the hidden representations enter GRU to generate the utterance level style embedding. The detailed process appears in equation \ref{eq:mixer}, where $MT$ is the six layers of the MixerTTS blocks. We denote the time step by $T$, the discrete code dimension by $D$ and the hidden dimension of the TTS model by $h$. The final output of the DC CoMix is utterance-level style embedding $s'$.
\begin{align}
  s^{h \times T} = MT({S}^{D\times{T}}) \nonumber \\
  s'^{h} = GRU(s^{h \times T})
\label{eq:mixer}
\end{align}

\subsection{DC CoMix TTS}
Discrete codes are not Mel scaled since the audio compression model is conditioned on the raw waveform. Thus, these are not suitable additional information for generating a Mel-spectrogram. Therefore, we choose the end-to-end TTS model as our backbone, which does not use intermediate features and directly synthesizes raw waveform from given texts. In fact, we observed a blurry Mel-spectrogram when the discrete code was used for additional information in the cascade model. 

The style embedding from the reference encoder is globally conditioned on the overall TTS system. As shown in Figure 1, the style embedding is fed into the posterior encoder, waveform decoder, stochastic duration predictor, and flow module. It is concatenated with each input in the corresponding module before entering it. It is the same schema used for speaker embedding in multi-speaker VITS. In addition to the style embedding, we keep the speaker look-up table to maintain speaker identity in the inference. As a result, we get a high speaker similarity with the desired speaker rather than the speaker of the reference speech. The experiment results are shown in \ref{sec:result}. The learning objective is the same as in the VITS. The decoder receives hidden representation segments from the posterior encoder following \cite{JimenezRezende2015VariationalIW}\cite{donahue2021endtoend}.
%
%
\section{Experiment and Results}

\subsection{Training details}
Nemo-toolkit \cite{kuchaiev2019nemo} was used for the implementation. We used IPA-G2P and EnglishPhonemeTokenizer from the Nemo-toolkit were applied for text preprocessing. The detailed configuration for text preprocessing is shown in the source code.\footnote{\url{https://github.com/lakahaga/dc-comix-tts}} The sampling rate was set to 24kHz to have the same time resolution in the audio compression model. For the linear spectrogram, we set FFT, window, and hop size to 1024, 1024, and 256, respectively. For VITS and MixerTTS blocks in the reference encoder, we used the same model configuration as in the examples from the Nemo-toolkit. The TTS system was trained using the AdamW \cite{Loshchilov2017DecoupledWD} optimizer with an initial learning of 2e-4, $\beta_1=0.9$ and $\beta_2=0.99$. The learning rate decayed by an exponential rate of 0.999875 for each epoch. Following VITS, we also used randomly extracted segments with a size of 32. Mixed precision and batch sampler were used to achieve faster training and 4 NVIDIA V100 32GB GPUs with a batch size of 80 for each GPU were employed. DC Comix TTS was trained until it reached 120k steps. 

\subsection{Experimental setup}
We trained models using the VCTK dataset \cite{veaux2017cstr}, a multi-speaker English dataset with various accents. We re-sampled the waveforms from 48kHz to 24kHz to match the time resolutions in Encodec. Samples shorter than 0.7 seconds were excluded. The training and test dataset was split at a ratio of 9:1 and ten samples from each speaker in the training dataset were set aside for validation. The train set and validation set comprises 38.5 hours and 1 hour, respectively. The test set consisted of the same speakers in the training set.

We used GST-extended FastSpeech2 with Hifi-GAN \cite{kong2020hifi} as the baseline. The open-source model from Espnet \cite{hayashi2020espnet} available from here\footnote{\url{https://zenodo.org/record/4036266\#.Y_w-N-xBy-Y}} was used. Both subjective and objective evaluations were applied to achieve a thorough analysis on the proposed method. For subjective evaluation, we evaluated our method using MOS evaluation. The ten native English speakers were asked to rate the naturalness and similarity of the audio samples on a 5-point scale, where 5 is the best score and 1 is the worst. We presented 30 audio samples from the speakers. 
For Similarity MOS (SMOS), the reference speech was randomly given from the test set, ensuring its content differed from the given text. The evaluators were asked to rate the acoustic similarity between the reference and synthesized speech, except for the timbre. This is because the desired speaker identity of the synthesized speech is not the one reference speech has. 

For the objective evaluation of the similarity between the two audio samples, we used speaker embedding and discrete code. The objective metrics are shown in Equation \ref{eq:obj-eval}.  Speaker similarity $cos(\theta_s)$ checks the preservation of speaker identity by cosine similarity between X-vectors from the ground truth $Spk_{GT}$ and the synthesized audio $Spk_{\hat{y}}$. The ground truth is the original speech of the given text and speaker identity. We use a pre-trained X-vector model \cite{snyder18_odyssey}.
In addition, discrete code similarity $cos(\theta_d)$ measures the acoustic similarity by cosine similarity of the discrete code of the reference speech $DC_{ref}$ and the one that of the synthesized speech $DC_{\hat{y}}$. We also used Encodec for evaluation. The average of cosine similarities of all 4397 samples in the test sets was calculated.

\begin{align}
    cos(\theta_s) = \frac{Spk_{GT} \cdot Spk_{\hat{y}}}{\lVert Spk_{GT}\rVert \lVert Spk_{\hat{y}}\rVert}, cos(\theta_d) = \frac{DC_{ref} \cdot DC_{\hat{y}}}{\lVert DC_{ref}\rVert \lVert DC_{\hat{y}}\rVert}
\label{eq:obj-eval}
\end{align}
\vspace{-3.5em}

\subsection{Results}
\label{sec:result}
Table \ref{tab:main-result} shows the overall evaluation result. The proposed model achieves a higher score in all evaluations than the baseline. In subjective evaluation, the score difference of SMOS is larger than that of MOS. 
The proposed model has more strength in acoustic similarity than in the naturalness of speech compared to the baseline. 
The score difference in MOS is small, but it is a promising result, considering the baseline uses more advantageous information to synthesize continuous speech signals. Regarding SMOS, the proposed model achieves better performance with a significant difference. 

\begin{table}[t]
\centering
\begin{tabular}{L{1.1cm}| C{1.46cm} | C{1.46cm} |  C{1cm} | C{1.1cm} }
 \toprule
 \multirow{2}{*}{Method} & \multicolumn{2}{c|}{Subjective} & \multicolumn{2}{c}{Objective} \\\cline{2-5}
    & MOS& SMOS & $cos(\theta_{s})$& $cos(\theta_{d})$  \\ \hline
    GT &4.53$\pm 0.05$ &-&-& -\\\hline
    Baseline& 3.68 $\pm 0.09$&3.13 $\pm 0.05$& 0.88&0.61\\
    Ours & \textbf{4.00} $\pm 0.05$&\textbf{3.98}$\pm 0.04$&\textbf{0.97}&\textbf{0.71}\\
\bottomrule
\end{tabular}

\caption{Performance of DC CoMix TTS compared to the baseline. MOS and SMOS with 95\% confidence for subjective evaluation, and Speaker Similarity cos($\theta_{s}$) and Acoustic Similarity cos($\theta_{d}$) for objective evaluation.}
\vspace{-2.5em}
\label{tab:main-result}
\end{table}

In the objective evaluation, our method also achieved a higher score than the baseline. Our system can retain the desired speaker identity while GST-FastSpeech2 cannot. Our system has separate modules for learning speaker identity and style information. We explicitly disentangle the speaker identity and speaker-independent prosody. On the other hand, GST-FastSpeech2 does not have a partitioned module for speaker identity. In the case of acoustic similarity, the proposed method also is superior to GST-FastSpeech2. Although GST-FastSpeech2 uses additional information at a higher resolution (80-bin), the proposed system achieves results that are more similar to the reference speech.  

\subsection{Ablation Study}

The main factors in DC CoMix TTS are the use of discrete code from the audio compression model and the reference encoder architecture. We analyzed how each component affects the performance. First, we compared our method with the architecture of GST at the same input representation (w/o Mixer). Then, we measured the differences between the spectrum-based features and the proposed input representation (w/o dc). We used a linear spectrogram instead of a Mel-spectrogram because the backbone VITS does not use the Mel-scale feature. We conducted both subjective and objective evaluations for the ablation study. Other than the differences above, the rest of the configuration and hyper-parameters were sustained. The results are shown in Table \ref{tab:ablation}. 
\begin{table}[ht]
\centering{
\begin{tabular}{L{0.9cm}| C{1.46cm} | C{1.46cm} |  C{1.05cm} | C{1.05cm} }
 \toprule
 \multirow{2}{0cm}{Method} & \multicolumn{2}{c|}{Subjective} & \multicolumn{2}{c}{Objective} \\\cline{2-5}
& MOS&SMOS&$cos(\theta_{s})$&$cos(\theta_{d})$\\ \midrule
Ours                 & \textbf{4.00}$\pm 0.05$    &  \textbf{3.98}$\pm 0.04$   &   \textbf{0.97}    & \textbf{0.71}  \\ \midrule
w/o Mixer                  & 3.92 $\pm 0.04$            &  3.86 $\pm 0.07$           &   \textbf{0.97}    &  0.70          \\ 
w/o dc          & 3.89 $\pm 0.08$            &  3.25 $\pm 0.06$           &    0.96            &  0.69          \\ \bottomrule
\end{tabular}}
\caption{The audio quality(MOS), similarity with the reference speech(SMOS, cos($\theta_{d}$)), and Speaker Similarity cos($\theta_{s}$) comparisons for each model variant. MOS and SMOS are measured with 95\% confidence.}
\vspace{-1.5em}
\label{tab:ablation}
\end{table}

The subjective evaluation result shows the input representation affects the performance the most. Using linear spectrogram as input representation can generate natural speech, but the similarity with the reference speech is inferior to the model using discrete code. The high score in naturalness originates from the high resolution of the input representation. Like the comparison between the baseline and the proposed method, employing the discrete code has improved the similarity more than the naturalness.
Using MixerTTS blocks has less impact than the discrete code. Still, it reduced the model parameter by around 0.4 million.

The objective evaluation results of the three model variants show little difference from each other. Given that the linear spectrogram has 513 channels and the discrete code has eight channels, using discrete code achieves comparable results even with low dimensional information. The difference in the number of model parameters decreases by twelve million with discrete code. Although the difference may appear marginal, MLP-Mixer offers advantages over CNN-based architectures. It can be parallelized, whereas the CNN-based architecture cannot. From this result, we prove that using the discrete code is not only effective in addressing the content-leakage problem but also reduces the complexity of prosody transfer TTS models.

\begin{table}[th]
\centering
\small{
\begin{tabular}{l|c}
\toprule
Method                     &  \# of parameters \\ \midrule
Ours                 &       \textbf{86.6M}       \\
w/o Mixer                  &           87.01M           \\
w/o discrete code          &           99.8M            \\ \bottomrule
\end{tabular}%
}
\caption{Comparison of the number of model parameters for ablation study. The goal is to minimize the number of parameters.}
\vspace{-3.5em}
\label{tab:params}
\end{table}

\section{Discussion}
In prosody transfer, the stress of words should be preserved even though the rhythm and intonation of their utterance change. However, in reality, the stress in words is fluid. For example, the word ``institute" is pronounced \textipa{["In.st9.tu:t]}.  However, in the synthesized audio, it is pronounced \textipa{[In.st9"tu:t]}. 
The stress moved from the front to the end. 
This is critical because stress is often used to identify the part of speech of the word. 
As a result, changing stress can change the meaning of the word. 
Thus the transferred prosody may change the meaning of the sentence. As such, there is a limitation in the prosody transfer method. Further research must find a line between reserving its stress and simultaneously transferring prosody from other speech.

\section{Conclusion}
This paper proposes a more effective prosody modeling for prosody transfer called DC CoMix TTS. This system uses discrete code from the audio compression model as the input representation of the reference encoder. The discrete code has rich acoustic information due to the learning objective of its original model. In addition, it has a lower dimension than Mel-spectrogram, reducing the training cost. Furthermore, we utilize MixerTTS blocks as the reference encoder, resulting in a more simplified architecture. The DC CoMixer TTS is trained in an end-to-end manner. We evaluate our system with both subjective and objective evaluation. We suggest a new objective metric that is suitable for calculating the acoustic similarity between two audio samples using discrete code. In both evaluations, we demonstrate that the proposed system achieves superior performance in transferring speaker-independent prosody.

\section{Acknowledgment}
This work was supported by Institute of Information \& communications Technology Planning \& Evaluation (IITP) grant funded by the Korea government(MSIT) (No.2022-0-00621,Development of artificial intelligence technology that provides dialog-based multi-modal explainability)

\bibliographystyle{IEEEtran}
\bibliography{mybib}

\end{document}